\documentstyle[aps,12pt]{revtex}
\begin{document}
\title{The preservation of the individuality of 3d atoms in a solid}
\author{R.J. Radwa\'{n}ski and Z.~Ropka}
\address{Center for Solid State Physics, \'{s}w. Filip 5, 31-150 Krak\'{o}w, Poland.}
\maketitle

\begin{abstract}
On basis of analysis of experimental results for more than 200 compounds
with 3d and 4f elements we conjecture that atoms with unfilled 3d and 4f
shells preserve much of their atomic properties, manifested by the discrete
electronic structure, even then when they become the part of a solid. As a
consequence, electronic and magnetic properties of the 3d-ion containing
compounds are strongly affected by the existence of the atomic-like
crystal-field fine electronic structure.



Receipt date: Phys.Rev.Lett. 30.11.1999
\end{abstract}

The discovery of the high-temperature superconductivity in the copper oxides
has revealed the enormous shortage of our general understanding of the
3d-ion compounds. For example, the insulating state of La$_2$CuO$_4$
contradicts the standard band-structure result that predicts it to be metal.$%
^1$ This dramatic breakdown of the ordinary band-energy theory has been
known already for years for 3d-ion monooxides and it was Sir N.F.Mott who
has realized it already 50 years ago (due to him the insulating 3d compounds
are presently known as Mott insulators). His problem why NiO, for instance,
having unfilled 3d shell is insulator is still not clear. Different
reparations of the conventional band-structure theory, that treats
d-electron states to form the continuum energy band do not lead to the
consistent picture for 3d-ion systems. On basis on analysis of more than 100
compounds containing 3d elements we strongly believe that the incorrect
treatment of the 3d atom is the reason for it. By this Letter we put a
conjecture that 3d elements even then when they become the part of a solid
preserve much of their atomic properties similarily to the understanding of
the 4f rare-earth systems.

By atomic properties we understand the existence of the discrete energy
spectrum (with the energy separation even below 1 meV) and by this
conjecture we say that this discrete atomic-like electronic structure is
preserved when the 3d atom becomes the part of a solid. This discrete
electronic structure is superimposed on the electronic structure of the
whole compound. This atomic-like structure predominantly determines the
electronic, magnetic and transport properties as well as their temperature
dependence. The explanation of electronic, magnetic and transport properties
of solids is the main goal of the solid-state physics. The possibility of
calculations of the temperature dependence of macroscopic physical
properties of a solid is the great advantage of the present atomic-based
model as it is possible for 4f systems.

Coming to details, our conjecture means that n d electrons in the unfilled
3d shell form, like in the atom, a highly-correlated electron system 3d$^n$
by means of intra-atomic interactions. This object, having own spin and
orbital momentum, interacts with the surrounding charge and spin. As an
example, let consider LaCoO$_3$. In accordance with its insulating state we
have ionic configurations: La$^{3+}$Co$^{3+}$O$_3^{2-}$. The La$^{3+}$ and O$%
^{2-}$ ions are largely inactive in the electronic and magnetic properties
having the full Xe-like and Ne-like configuration though they are taking
important role in the bonding (via conventional electrostatic interactions,
also multipolar) and the formation of the solid due to the electronic
charges. Then we have to consider the discrete energy spectrum of the Co$%
^{3+}$ ion (3d$^6$ electronic system) that by intra-atomic interactions
takes the term $^5$D as the ground term (two Hund's rules). This ground term
being 25-fold degenerated is split by (multipolar) electrostatic
interactions (Stark-like effect) and in case of magnetically-ordered
systems, not the case of LaCoO$_3$, by Zeeman-like effect. Of course, for
the detailed discussion we need to know the exact charge and spin
distribution (the main goal of local-density approximation theories). It is
still the unsolved problem in the solid-state physics. There is, however, a
solution of this tremendous problem. We can learn a lot about the splitting
and the ground state already from the local symmetry of the multipole
electrostatic interactions (the crystal electrostatic field) with the use of
the group theory. In refs 2 and 3 we have shown that in the case of the
perovskite structure relevant to LaCoO$_3$ the ground state of the Co$^{3+}$
ion in the slightly-distorted octahedral symmetry is a nonmagnetic singlet. 
For getting of this singlet taking into account the intra-atomic spin-orbit
coupling is essential. This atomic-like explanation of the 50-years lasting
problem of the low-temperature diamagnetism and its temperature
transformation to the paramagnetic state deserves on the special attention
due to its physical simplicity and the good foundation within physical
concepts. Moreover, this our solution leads to the substantial unification
of atomic and solid-state physics.

The scientific intriguity of LaCoO$_3$ is related with the experimental fact
of the drastic violation of the theoretical S=2 temperature dependence of
the paramagnetic susceptibility at low temperatures, i.e. with the violation
of the Curie-Weiss law. According to us this drastic violation of the
Curie-Weiss law  is related with the existence of the discrete energy
states. In case of LaCoO$_3$ there are 15 discrete states [2,3] within 300
meV (5 times value of the spin-orbit coupling). It yields the average energy
separation of 20 meV. Obviously, the existence of such atomic-like fine
electronic structure affects enormously electronic and magnetic properties
of the whole compound. For the Mn$^{3+}$ ion in the octahedral
nearest-neighbour oxygen surrounding 10 states of the $^5$E$_g$ subterm are
contained within 10 meV [3]. The extremally small energy separation occurs
for the 3d$^n$ systems with the odd number of d electrons. For these systems
the Kramers doublet is the charge-formed ground state. This Kramers
degeneracy is later spontaneously removed by interactions involving the
spin, i.e. with the breaking the time-reversal symmetry and the formation of
the magnetic state.

We can ask: ''Is this atomic idea a new one in the solid-state physics?''
Yes and no. No, as most of experimentalists naturally discuss their results
in terms of local properties. Yes, as according to our knowledge noone has
been able to resist to presently-in-fashion solid-state physics theories
that simply ignore the existence of the atom in the solid arguing that the
solid is so many-body object and that there are so strong intersite
correlations that the individuality of atoms is lost. In the standard
band-structure calculations the d electrons are taken as itinerant forming a
band of about 5 eV width. In the band there is a continuum of the energy
states within 5 eV. No, as there are some text books written about the
crystal field, let mention a book of Abragam and Bleaney or Ballhausen [4].
Yes, as they applied the CEF\ approach to some diluted systems, not to the
concentrated ones. Yes, as they have not been consequent enough and by
discussing different crystal-field approaches (weak, strong, ...) with
further concepts they largely washed up the original idea. Please note that
in the strong CEF\ approach the n 3d electrons are treated as largely
independent, i.e. they do not form the highly-correlated 3d$^n$ system in
contrary to the present model. Our approach corresponds to the weak
crystal-field approach but we point out the fundamental importance of the
intra-atomic spin-orbit coupling, despite of its relative weakness for the
3d ions [5]. Also yes, as at present this atomic-like picture is enormously
prohibited in the leading physical journals and a little is said about the
discrete states in the magnetic and strongly-correlated electron system
conferences. We understood our approach as the continuation of (some) ideas
of Bethe, Kramers, Van Vleck, Mott, Anderson, Hubbard, ....who pointed out
the importance of local-scale effects on magnetic and electronic properties.

The existence of these discrete states, with the energy separation below 1
meV has been already well evidenced for conventional rare-earth systems (4f
shell) though strong discussion is going on about their existence in
anomalous rare-earth systems, namely with Ce and Yb. The good description
[6,7] of electronic and magnetic properties of intermetallics ErNi$_5$ and
NdNi$_5$ indicates on the substantial preservation of the individuality of
the Er$^{3+}$ and Ni$^{3+}$ ions also in intermetallic compounds. For
description of the intermetallics the atomic-like approach has been extended
to the individualized-electron (IND-E) model that considers the coexistence
of (4f $^n$)localized- and itinerant-electronic subsystems. Here we should
mention a surprisingly good description within the IND-E model of magnetic
and electronic properties of the actinides intermetallics UPd$_2$Al$_3,$ UGa$%
_2$ and NpGa$_2$ [8,9] as originating from the discrete states of the U$^{3+}
$/Np$^{3+}$ ions. Macroscopic properties can be obtained at the quantitative
level from the microscopic ones by the direct multiplification by the number
of the transition-metal ions involved (to be specific, the molar heat
capacity value, for example, is obtained from the atomic value via Avogadro
number). It is worth noting that our approach reveals very strong
correlation between the local magnetism, the local crystallographic symmetry
(Jahn-Teller effect) and the fine electronic structure.

We would like to add, preceding unfounded critics, that we do not claim that
everything can be explained only by single atoms but our point is that the
proper, i.e. physically adequate starting point for the discussion of
properties of the solid containing the open shell elements is the
consideration of its atomic states. Our numerous computer experiments point
out that e.g. the orbital moment has to be unquenched in the solid-state
physics of 3d-ion containing compounds and our approach enables it. For
instance, we have derived the orbital moment in NiO to be 0.54 $\mu _B$ what
amounts to 20 \% of the total moment (2.53 $\mu _B$) [10]. Moreover, one
should not consider our approach as the treatment of an isolated atom - we
clearly discuss the kation octahedra. The perovskite structure, for
instance, is built up from the corner sharing kation octahedra.

In conclusion, on basis of the extended analysis of experimental results for
the great number of compounds (more than 200, including 4f and 5f compounds)
we put a conjecture that in compounds containing atoms with unfilled
3d/4f/5f shells there exists the discrete electronic structure associated
with quasi-atomic states of the involved ions. The existence of such the
structure causes dramatic changes of the low-temperature electronic and
magnetic properties like the formation (or not) of the local magnetic moment
and its long-range magnetic order, temperature dependence of the magnetic
suceptibility and of the heat capacity [2,6-9].

The note added during the referee process (7.02.2000). For the better
illustration of our point of view the reader is asked to look into recent
Phys.Rev.Lett. papers. In Ref. 11 authors, considering states of two 3d
electrons of the V$^{3+}$ ion in V$_2$O$_3$, came out with the continuum
electronic structure spread over 2.5 eV (Figs 2 and 3). In Ref. 12 the
continuum electronic structure for six 3d electrons in FeO spreads over 8 eV
(Fig. 8). By this Letter we put the conjecture that in both cases there are
crystal-field discrete energy states as have been discussed in Refs 2 and 3
following the original idea of Bethe from 1929. In FeO, in the paramagnetic
state, quite similar structure to that presented in Refs 2 and 3 is realized.

\end{document}